# Anomalous violation of the OZI-rule in the $N\bar{N} \to \Phi\Phi$, $N\bar{N} \to \Phi\gamma$ reactions and instantons

N.I.Kochelev

Dipartimento di Fisica, Università di Pisa and INFN
Sezione di Pisa, I-56100 Pisa, Italy
and
Bogoliubov Laboratory of Theoretical Physics,
Joint Institute for Nuclear Research,
Dubna, Moscow region, 141980, Russia [1]

**Abstract**

It is shown, that specific properties of the instanton induced interaction between quarks leads to the anomalous violation of the OZI-rule in the $N\bar{N} \to \Phi\Phi$, $N\bar{N} \to \Phi\gamma$ reactions. In the framework of instanton model of the QCD vacuum, the energy dependence of the cross sections of these reactions is calculated.

---

[1]Permanent address; E-mail:kochelev@thsun1.jinr.dubna.su

The OZI rule [1] is successfully used to explain suppression of the processes which include disconnected quark lines. However, recently, anomalous violation of this rule in some channels with creation of the $\Phi$-mesons was found in the $N\bar{N}$ annihilation (see review [2] and referencies therein). The most significant diviations from OZI rule have been observed in the reactions $N\bar{N} \to \Phi\Phi$ [3] and $N\bar{N} \to \Phi\gamma$ [4], where measured values of the cross sections differ from predicted ones on the several orders.

Some explainations of this puzzle were suggested. So, in [5], the large mixture of the strange quarks inside nucleon has been supposed. The more conventional approach, based on the rescattering, was also involved to explain these results [6].

Recently, the new mechanism of the OZI-rule violation in the $N\bar{N}$ annihilation has been proposed [7]. This mechanism is based on the specific interaction between quarks [8] which is induced by the instantons [9]. It has been shown, that taking into account of the instanton induced interaction, allow us to explain, at least qualitatively, some features of the OZI rule violation in $N\bar{N}$ annihilation. It should be mention also, that at the begining this interaction has been used to solve the several old problems of the hadron spectroscopy, in particularly, to solve the problem of the large violation of the OZI rule in the wave functions of the pseudoscalar nonet mesons [10]. At the present time, the same interaction is considered as the evident QCD mechanism to explain so-called "spin crisis" which is connected with the large OZI rule violations in the spin-dependent structure functions [11].

The main goal of this article is the calculation of the instanton contribution to the cross sections of the processes $N\bar{N} \to \Phi\Phi$ and $N\bar{N} \to \Phi\gamma$.

The effective Lagrangian for the interaction induced by the instantons for the number of the flavors $N_f = 3$ and for massless quarks has the following form [12]:

$$\begin{aligned}
\mathcal{L}_{eff}^{(N_f=3)} &= \int \frac{d\rho}{\rho^5} d(\rho) \left\{ \prod_{i=u,d,s} (-\frac{4\pi^2}{3}\rho^3 \bar{q}_{iR} q_{iL}) + \right. \\
&\quad \frac{3}{32}(\frac{4}{3}\pi^2\rho^3)^2 [(j_u^a j_d^a - \frac{3}{4} j_{u\mu\nu}^a j_{d\mu\nu}^a)(-\frac{4}{3}\pi^2\rho^3 \bar{q}_{sR} q_{sL}) + \\
&\quad \frac{9}{40}(\frac{4}{3}\pi^2\rho^3)^2 d^{abc} j_{u\mu\nu}^a j_{d\mu\nu}^b j_s^c + 2perm.] + \frac{9}{320}(\frac{4}{3}\pi^2\rho^3)^3 d^{abc} j_u^a j_d^b j_s^c + \\
&\quad \left. \frac{ig f^{abc}}{256}(\frac{4}{3}\pi^2\rho^3)^3 j_{u\mu\nu}^a j_{d\nu\lambda}^b j_{s\lambda\mu}^c + (R \longleftrightarrow L) \right\},
\end{aligned} \qquad (1)$$

where $q_{R,L} = \frac{(1\pm\gamma_5)}{2}q(x)$, $j_i^a = \bar{q}_{iR}\lambda^a q_{iL}$, $j_{i\mu\nu}^a = \bar{q}_{iR}\sigma_{\mu\nu}\lambda^a q_{iL}$, and $d(\rho)$ is the instanton density and $\rho$ is their size. It should be point out that Lagrangian (1) has been obtained considering quark zero modes in the instanton field. In this case the point-like interaction (1) appears in the limit of small size of the instantons $k_i\rho \to 0$, where $k_i$ are momenta of the incoming and outcoming quarks. Taking into account the finite size of the instanton leads to some form factor in (1) which is proportional to the Fourier transformation of the quark zero modes [13]. In the regular gauge it is [14]

$$f(k_i) = exp(-\rho \sum_i k_i). \tag{2}$$

The density of the instantons for $N_f = 3$ and $N_c = 3$ is given by [15]

$$d(\rho) = 3.64 \cdot 10^{-3} (\frac{2\pi}{\alpha_s(\rho)})^6 exp[-\frac{2\pi}{\alpha_s(\rho)} S(\rho)], \tag{3}$$

where

$$S(\rho) = 1 - \frac{\pi^3}{16\alpha_s(\rho)} < 0 \mid \frac{\alpha_s}{\pi} G_{\mu\nu}^2 \mid 0 > \rho^4. \tag{4}$$

The second term in the equation (4) takes into account interactions between instantons [16]. This term leads to the anomalous growth of the density of the instantons with the increasing of their size. Anomalous growth of the density is one of the main problems of practically all the calculations of the contributions of the instantons to observelable values, since the necessity to introduce some additional mechanism of density stabilization is appeared (see discussion in [17]). Recently, in [18] it was shown that unitarity condition leads to a restriction on the magnitude of the contribution of the interaction between instantons into the equation (4) by the value 1/2. Therefore, in our calculation below, we will use the following approximation for $S(\rho)$

$$S(\rho) = \begin{cases} 1 - \frac{\pi^3}{16\alpha_s(\rho)} < 0 \mid \frac{\alpha_s}{\pi} G_{\mu\nu}^2 \mid 0 > \rho^4, & \text{if } \rho < \rho_0; \\ 0.5 & \text{if } \rho \geq \rho_0. \end{cases}, \tag{5}$$

where

$$\rho_0 = (\frac{8\alpha_s}{\pi^3 < 0 \mid \frac{\alpha_s}{\pi} G^2 \mid 0 >})^{\frac{1}{4}}, \tag{6}$$

and $< 0 \mid \frac{\alpha_s}{\pi} G^2 \mid 0 > = (330 Mev)^4$ is the gluon condensate. It should be mentioned, that a similar approximation was used to estimate contribution of the instanton to DIS structure functions in the paper [19].

Contribution of the instanton to the $N\bar{N} \to \Phi\gamma$ and $N\bar{N} \to \Phi\Phi$ reactions is presented on the diagrames in the Fig.1. The part of the Lagrangian (1) which has the right quantum numbers and can contribute to the cross sections of these processes is

$$\Delta \mathcal{L}_{eff}^{(N_f=3)} = \int \frac{d\rho}{\rho^5} d(\rho) \frac{<0 \mid \bar{q}q \mid 0>}{4} f(k_i)(\frac{4\pi^2}{3}\rho^3)^3((\bar{u}i\gamma_5 u + \bar{d}i\gamma_5 d)(\bar{s}i\gamma_5 s)). \quad (7)$$

In the formula (7) the appearence of the quark condensate $<0 \mid \bar{q}q \mid 0>$ follows from the contraction of the two quark legs from six-fermionic vertex. Form factor $f(k_i)$ is determined by the energy of the colliding particles and equals

$$f(k_i) = e^{-2\rho\sqrt{s}}. \quad (8)$$

The matrix elements from Lagrangian (7) for the reactions $N\bar{N} \to \Phi\gamma$, $N\bar{N} \to \Phi\Phi$ are

$$M(N\bar{N} \to \Phi\gamma) = \lambda_{inst}(s) < N\bar{N} \mid (\bar{u}i\gamma_5 u + \bar{d}i\gamma_5 d)\bar{s}i\gamma_5 s \mid \Phi\gamma>, \quad (9)$$

$$M(N\bar{N} \to \Phi\Phi) = \lambda_{inst}(s) < N\bar{N} \mid (\bar{u}i\gamma_5 u + \bar{d}i\gamma_5 d)\bar{s}i\gamma_5 s \mid \Phi\Phi>. \quad (10)$$

where

$$\lambda_{inst}(s) = \int \frac{d\rho}{\rho^5} d(\rho) \frac{<0 \mid \bar{q}q \mid 0>}{4} (\frac{4\pi^2}{3}\rho^3)^3 e^{-2\rho\sqrt{s}}. \quad (11)$$

We will use the axial anomaly and vector dominance model (VDM) to calculate the matrix elements (9), (10). So one can reduce the equations (9), (10) to the product of the matrix elements

$$< N\bar{N} \mid (\bar{u}i\gamma_5 u + \bar{d}i\gamma_5 d)\bar{s}i\gamma_5 s \mid \Phi\gamma > = \quad (12)$$

$$< N\bar{N} \mid (\bar{u}i\gamma_5 u + \bar{d}i\gamma_5 d) \mid 0 > < 0 \mid \bar{s}i\gamma_5 s \mid \Phi\gamma >, \quad (13)$$

$$< N\bar{N} \mid (\bar{u}i\gamma_5 u + \bar{d}i\gamma_5 d)\bar{s}i\gamma_5 s \mid \Phi\Phi > = \quad (14)$$

$$< N\bar{N} \mid (\bar{u}i\gamma_5 u + \bar{d}i\gamma_5 d) \mid 0 > < 0 \mid \bar{s}i\gamma_5 s \mid \Phi\Phi >, \quad (15)$$

Some of these matrix elements can be determined by means of the VDM connection [20]:

$$< 0 \mid \bar{s}i\gamma_5 s \mid \Phi\gamma > = \frac{g_\Phi}{e} < 0 \mid \bar{s}i\gamma_5 s \mid \gamma\gamma >, \quad (16)$$

$$< 0 \mid \bar{s}i\gamma_5 s \mid \Phi\Phi > = (\frac{g_\Phi}{e})^2 < 0 \mid \bar{s}i\gamma_5 s \mid \gamma\gamma >, \quad (17)$$

and the relation given by axial anomaly (see for example [21])

$$< 0 \mid \bar{s}i\gamma_5 s \mid \gamma\gamma > = \frac{1}{m_s^*} \frac{\alpha_s}{2\pi} N_c e_s^2 F_{\mu\nu}^{(1)} \tilde{F}_{\mu\nu}^{(2)}. \quad (18)$$

In the equations (16), (17), (18) $g_\Phi^2/4\pi = 13.3$ is $\Phi$-meson coupling constant with photon [20],

$$F^{(1)}_{\mu\nu} = k^{(1)}_\mu \epsilon^{(1)}_\nu - k^{(1)}_\nu \epsilon^{(1)}_\mu, \tag{19}$$

$$\widetilde{F}^{(2)}_{\mu\nu} = \frac{1}{2}\epsilon^{\mu\nu\rho\sigma}(k^{(2)}_\mu \epsilon^{(2)}_\nu - k^{(2)}_\nu \epsilon^{(2)}_\mu), \tag{20}$$

$k^{(i)}_\mu (\epsilon^{(i)}_\mu)$ is the momentum (polarization) of the photons ($\Phi$-mesons),

$$m_s^* = m_s^{cur} + m_q^*, \quad m_q^* = -\frac{2}{3}\pi^2 \rho_c^2 < 0 \mid \bar{q}q \mid 0 > \tag{21}$$

is the mass of the strange quark ($m_s^*$, $m_s^{cur} = 150 MeV$) and the masses of the non-strange quarks ($m_u^* = m_d^* = m_q^*$), $\rho_c = 1.6 GeV^{-1}$ is average size of the instanton in the QCD vacuum [14].

One can estimate the remaining matrix element $< N\bar{N} \mid (\bar{u}i\gamma_5 u + \bar{d}i\gamma_5 d) \mid 0 >$ by using the relation given by gluon axial anomaly:

$$< N\bar{N} \mid (\bar{u}i\gamma_5 u + \bar{d}i\gamma_5 d) \mid 0 > = -\frac{1}{2m_q^*} < N\bar{N} \mid \frac{\alpha_s}{4\pi}G_{\mu\nu}\widetilde{G}_{\mu\nu} \mid 0 > \tag{22}$$

$$= -\frac{M_N}{m_q^*}g_A^0 \bar{N}(p_1)i\gamma_5 \widetilde{N}(p_2). \tag{23}$$

To get equation (23), the connection between matrix element of topological charge density and singlet axial vector coupling constant which determines of determines of the part of the proton spin carried by quarks, was used [22]. In the paper [23] from analysis of the experimental data on spin-dependent structure functions [24] the value of the $g_A^0$ was extracted

$$g_A^0 = 0.31 \pm 0.07. \tag{24}$$

By taking into account all factors, we get the final formulas for cross sections:

$$\sigma_{N\bar{N}\to\Phi\Phi}(S) = \frac{0.39}{8\pi}(\frac{\lambda_{inst}(S)M_N g_A^0}{12 m_s^* m_q^* \pi}(\frac{g_\Phi^2}{4\pi}))^2 \cdot$$

$$\frac{((S - 2m_\Phi^2)^2 - 4m_\Phi^4)(S - 4m_\Phi^2)^{\frac{1}{2}}}{(S - 4M_N^2)^{\frac{1}{2}}} \ (mb) \tag{25}$$

and

$$\sigma_{N\bar{N}\to\Phi\gamma}(S) = \frac{0.39}{4\pi}\left(\frac{\lambda_{inst}(S)M_N g_A^0}{12m_s^* m_q^* \pi}\right)^2 \alpha\left(\frac{g_\Phi^2}{4\pi}\right) \cdot$$
$$\frac{(S-m_\Phi^2)^3}{(S-4M_N^2)^{\frac{1}{2}} S^{\frac{1}{2}}} \quad (mb). \tag{26}$$

For numerical calculation we used NLO approximation for strong coupling constant:

$$\alpha_s(\rho) = -\frac{2\pi}{\beta_1 t}\left(1 + \frac{2\beta_2 \log t}{\beta_1 t}\right), \tag{27}$$

where

$$\beta_1 = -\frac{33-2N_f}{6}, \quad \beta_2 = -\frac{153-19N_f}{12} \tag{28}$$

and

$$t = \log\left(\frac{1}{\rho^2 \Lambda^2} + \delta\right). \tag{29}$$

In equation (29) the parameter $\delta \approx 1/\rho_c^2 \Lambda^2$ provides a smooth interpolation the value of the $\alpha_s(\rho)$ from perturbative ($\rho \to 0$) to the nonperturbative region ($\rho \to \infty$) [25].

Result of the calculation of the cross section of the process $N\bar{N} \to \Phi\Phi$ is presented in Fig.2 together with the preliminary experimental data from JETSET Collaboration [4], for energies from threshold up to $\sqrt{S} = 2.4 GeV$, and the data of the R-704 Collaboration at $\sqrt{S} = 3 GeV$ [26]. There is a good agreement between model and experimental data. The small discrepance, probably, can be connected with uncertainties in the total normalization of the value of the cross sections in the JETSET Collaboration (see [27]). Instanton model correctly discribes observable strong energy dependence of the violation of the OZI rule. This dependence is due, in fact, to the strong sensibility of the value of the integral (11) the value of the initial energy $\sqrt{S}$, through the instanton density. Therefore, precise measurement of this dependence can give the very important information on the distribution of the instantons in the QCD vacuum.

It should be stressed, that OZI rule predicts the value of the cross section $\sigma^{OZI} \approx 10 nb$ which does not depend on the value of energy and is about two order smaller then the experimental data.

In Fig.3 the prediction of the cross section $\sigma(N\bar{N} \to \Phi\gamma)(S)$ is given. By using the experimental interpolation of the data on the total cross section of the $N\bar{N}$ annihilation

[28]
$$\sigma_{N\bar{N}}^{total} = -30.1 + \frac{46.6}{p} + 60.3p \ (nb), \tag{30}$$

where $p$ is the antiproton momentum, one can estimate the branching ratio of the $p\bar{p} \to \Phi\gamma$ reaction at rest

$$Br(p\bar{p} \to \Phi\gamma) = 1.8 \cdot 10^{-5}. \tag{31}$$

The value (31) is in very good agreement with the Crystal Barrel Collaboration result [3]

$$Br(p\bar{p} \to \Phi\gamma) = (1.7 \pm 0.4) \cdot 10^{-5}. \tag{32}$$

The OZI rule contradics these experimental data. So, if one uses the value of the branching ratio for the reaction $p\bar{p} \to \omega\gamma$

$$Br(p\bar{p} \to \omega\gamma) = (6.8 \pm 1.8) \cdot 10^{-5}, \tag{33}$$

given by the same collaboration [29], then this rule predicts

$$Br(p\bar{p} \to \Phi\gamma) = tan^2\Theta Br(p\bar{p} \to \omega\gamma) = 2.8 \cdot 10^{-7}(10^{-8}) \tag{34}$$

for the quadratic (linear) Gell-Mann-Okubo mass formula [2].

Thus, taking into account the instanton induced interaction between quarks allows us to explain the anomalous violation of the OZI rule in the reaction $p\bar{p} \to \Phi\Phi$, $p\bar{p} \to \Phi\gamma$.

We can conclude that significant violation of the OZI rule, which was revealed in the $N\bar{N}$ annihilation, is the nontrivial manifestation of the complex structure of the QCD vacuum connected with existence of the instantons.

Therefore, the investigation of the violation of the OZI rule in the $N\bar{N}$ annihilation can allow to obtain the very useful information on the structure of the ground state of the theory of the strong interaction, the QCD vacuum.

The author is sincerely thankful P.N.Bogolubov, A.E.Dorokhov, A.Fässler, M.Fässler, H.Petry, E.Klempt, D.Kharzeev, S.B.Gerasimov, B.Mench, M. Mintchev, M.G.Sapozhnikov, Yu.A.Simonov, B-S.Zou, for useful discussions, Prof. A. Di Giacomo for warm hospitality at Università di Pisa and INFN for support.

---

[2] In the equation (34) $\Theta$ is the deviation from the ideal mixing angle ($35.3^0$).

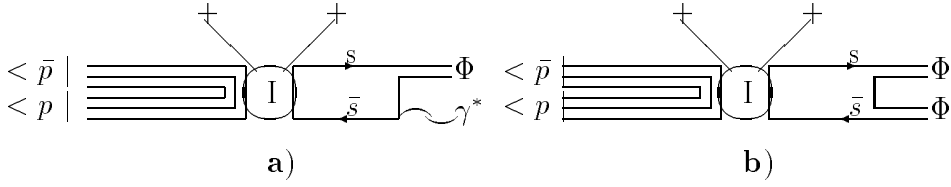

**Fig.1** Contribution of the instantons to the reactions a) $N\bar{N} \to \Phi\gamma$ and b) $N\bar{N} \to \Phi\Phi$. The instanton is denoted as I and crosses mark the quark legs which are connected through vacuum.

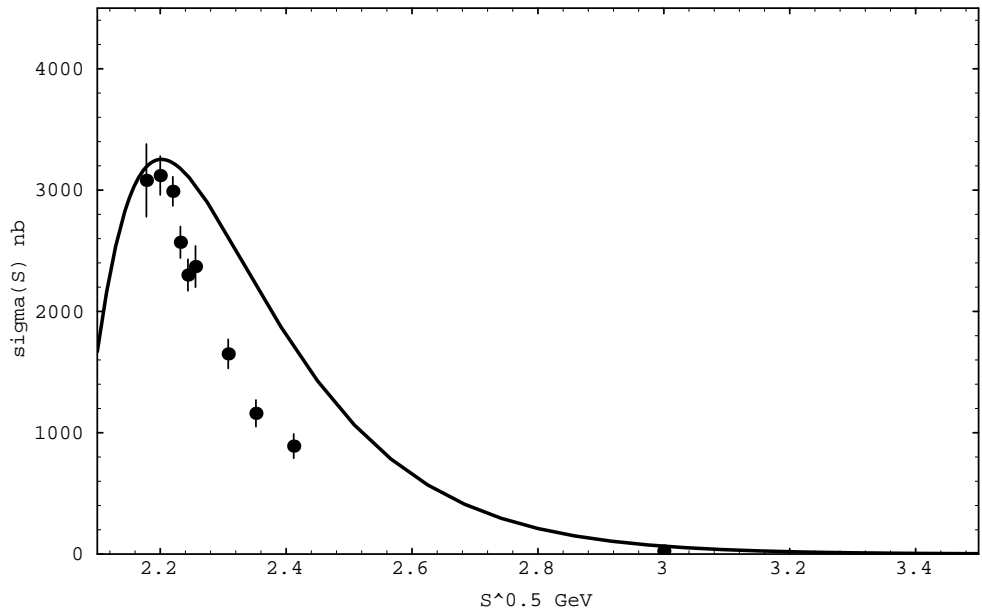

**Fig.2** Cross section of the reaction $N\bar{N} \to \Phi\Phi$. Solid line is the prediction of the instanton model and experimental data are results of the JETSET and R-704 Collaborations.

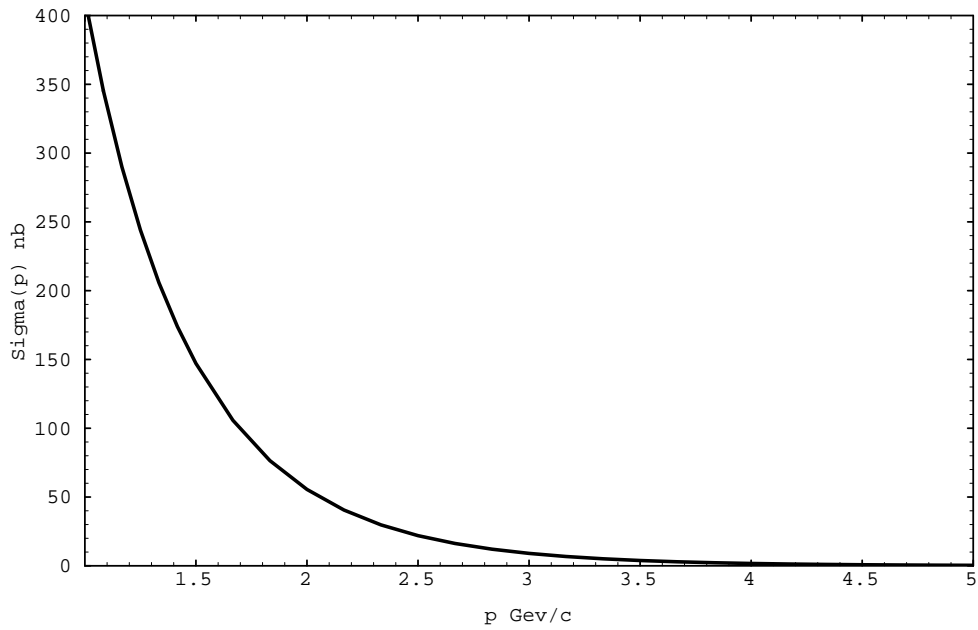

**Fig.3** Cross section of the reaction $N\bar{N} \to \Phi\gamma$. Solid line is the prediction of the instanton model.